\documentclass[a4paper, conference]{IEEEtran}

\usepackage{cite}
\usepackage{amsmath,amssymb,amsfonts}
\usepackage{algorithm,algorithmic}
\usepackage{graphicx}
\usepackage{textcomp}
\usepackage{geometry}
\usepackage{url}
\usepackage[caption=false,font=footnotesize]{subfig}
\def\BibTeX{{\rm B\kern-.05em{\sc i\kern-.025em b}\kern-.08em
		T\kern-.1667em\lower.7ex\hbox{E}\kern-.125emX}}

\begin{document}

\title{Multi-UAV Collaborative Trajectory Planning for Seamless Data Collection and Transmission}

\author{\IEEEauthorblockN{Rui Wang\textsuperscript{1}, Kaitao Meng\textsuperscript{2}, and Deshi Li\textsuperscript{1}}
\IEEEauthorblockA{\textsuperscript{1}Electronic Information School, Wuhan University, Wuhan, China.\\
\textsuperscript{2}Department of Electronic and Electrical Engineering, University College London, UK.\\
Emails: \textsuperscript{1}\{ruiwang, dsli\}@whu.edu.cn, \textsuperscript{2}kaitao.meng@ucl.ac.uk}
}

\maketitle

\begin{abstract}
Unmanned aerial vehicles (UAVs) have attracted plenty of attention due to their high flexibility and enhanced communication ability. 
However, the limited coverage and energy of UAVs make it difficult to provide timely wireless service for large-scale sensor networks, which also exist in multiple UAVs.
To this end, the advanced collaboration mechanism of UAVs urgently needs to be designed.
In this paper, we propose a multi-UAV collaborative scheme for seamless data collection and transmission, where UAVs are dispatched to collection points (CPs) to collect and transmit the time-critical data to the ground base station (BS) simultaneously through the cooperative backhaul link.
Specifically, the mission completion time is minimized by optimizing the trajectories, task allocation, collection time scheduling, and transmission topology of UAVs while ensuring backhaul link to the BS.
However, the formulated problem is non-convex and challenging to solve directly.
To tackle this problem, the CP locations and transmission topology of UAVs are obtained by sensor node (SN) clustering and region division.
Next, the transmission connectivity condition between UAVs is derived to facilitate the trajectory discretization and thus reduce the dimensions of variables. 
This simplifies the problem to optimizing the UAV hovering locations, hovering time, and CP serving sequence.
Then, we propose a point-matching-based trajectory planning algorithm to solve the problem efficiently.
The simulation results show that the proposed scheme achieves significant performance gains over the two benchmarks.

\end{abstract}


\section{Introduction}
The potential applications in future 6G networks rely on real-time collection and transmission of massive sensory data, thus facilitating the era of intelligent connection of everything \cite{Ferrag2023edge}.
Driven by high mobility and flexible deployment, unmanned aerial vehicles (UAVs) can act as aerial base stations (BSs) or sensors to provide enhanced wireless connectivity and sensing services \cite{meng2023multi}, which are applied in wide scenarios, such as environment monitoring, search-and-rescue, aerial inspection, etc  \cite{Wang2024Rechargeble}.
Nonetheless, due to the limited coverage and energy, a single UAV typically requires a lot of time and power resources to perform sensing and communication tasks, especially when sensor nodes (SNs) are distributed over a large area \cite{wei2022uav}.
Therefore, the collaboration of multiple UAVs to provide efficient communication services for large-scale networks becomes a promising solution \cite{Wu2018Joint}.

Typical multi-UAV-assisted data collection mainly focuses on task-level collaboration \cite{Xu2022Deep}, where the non-overlapping tasks are allocated to different UAVs to improve data acquisition efficiency. 
For instance, in \cite{Xu2022Deep}, multiple UAVs are employed to cooperatively perform sensing tasks in parallel, thereby reducing the mission time. 
However, this task allocation overlooks the cooperative strategy for the seamless transmission of collected data to the BS, which may affect the timeliness of sensory data.
Most recently, several works study the deployment or trajectory optimization issue for fresh data collection in multi-UAV-assisted Internet-of-Things (IoT) systems \cite{Sabzehali2022optimizing}, \cite{fu2024collaborative}.
For example, the authors in \cite{fu2024collaborative} employed an improved adaptive large neighborhood search algorithm for path planning and relay position determination of UAVs to achieve low Age of Information (AoI) data collection.
Nonetheless, in the above works, UAVs are typically deployed statically or conduct data transmission at fixed positions, which may incur huge costs of massive UAVs or extra time to fly to the designated locations.
In addition, full-time connectivity maintenance during UAV movement is important to enable cooperative teamwork and seamless communication among network entities \cite{Yanmaz2024}.
Without the backhaul connectivity link, the isolated UAVs cannot transmit the collected data back in real-time, resulting in mission inefficiency or failure \cite{Albu-Salih2018}.
On the other hand, it is also desirable for the controller since it allows the BS to keep track of the mission status and conduct safe operations of UAVs \cite{She2019UltraReliableAL}.
Therefore, connectivity maintenance is essential in multi-UAV collaborative networks for real-time and reliable data collection and transmission.

To ensure seamless data transmission through the cooperative backhaul link, UAVs perform both the collection and relay tasks, simultaneously fly to SNs, and promptly collect and transmit data.
However, several conflicting objectives must be carefully balanced when minimizing the overall mission completion time:
Firstly, for data collection, the task allocation typically aims at minimizing the total path length of UAVs. 
However, this causes some distant UAVs to require extended flight times to BS for data offloading due to the lack of a backhaul link, which increases the flight time and affects the timeliness of data.
Secondly, for data transmission, UAVs need to maintain a cooperative backhaul connection to the BS during the whole flight, which makes the trajectory of the UAV affected by the locations of the allocated SNs and also the locations of other UAVs.
Thirdly, to reduce the total time, the SN collection scheduling of each UAV needs to be carefully designed to obtain shorter collection time and flight time.
Therefore, it is highly complex to acquire the optimal cooperative trajectories of UAVs.

With the above consideration, we propose a multi-UAV collaborative scheme for seamless data collection and transmission, where UAVs are dispatched to collection points (CPs) to collect data and transmit to BS simultaneously through cooperative backhaul links.
In addition to collaboration on the collection, our proposed scheme also collaborates on data transmission, forming a seamless transmission link to a distant BS.
Specifically, a mission completion time minimization problem is formulated by optimizing the trajectories of UAVs, collection task allocation, collection time scheduling, and transmission topology of UAVs while guaranteeing backhaul connectivity to the ground BS in the whole mission.
However, solving this optimization problem is highly non-trivial since it is non-convex and involves infinite variables with closely coupled trajectories of UAVs.
To address this issue, the problem is first simplified to the joint optimization of UAV hovering locations, hovering time, and CP serving sequence.
Then, we propose a point-matching-based trajectory planning algorithm to minimize the completion time efficiently.
The main contributions are summarized as follows:
\begin{itemize}
    \item We propose a multi-UAV collaborative scheme for seamless data collection and transmission, and formulate a completion time minimization problem by optimizing the trajectories, task allocation, collecting time scheduling, and transmission topology of UAVs.
    \item We derive the transmission connectivity condition between UAVs, which guides the trajectory discretization to reduce the dimensions of variables.
    Then the problem can be transformed to optimize UAV hovering locations, hovering time, and CP serving sequence.
    \item We propose a point-matching-based trajectory planning algorithm, where the CPs are matched for simultaneous collection to reduce the total hovering time. 
    Then the lower bound path is referred to optimize the trajectories of UAVs to reduce the total flight time.
\end{itemize}

The rest of the paper is organized as follows. 
Section \uppercase\expandafter{\romannumeral2} presents the system model and problem formulation. 
Section \uppercase\expandafter{\romannumeral3} develops the trajectory planning algorithm. 
Section \uppercase\expandafter{\romannumeral4} provides simulation results. 
Section \uppercase\expandafter{\romannumeral5} concludes this paper.

\section{System Model and Problem Formulation}
\begin{figure}[t]
    \centering{\includegraphics[width=3.1in]{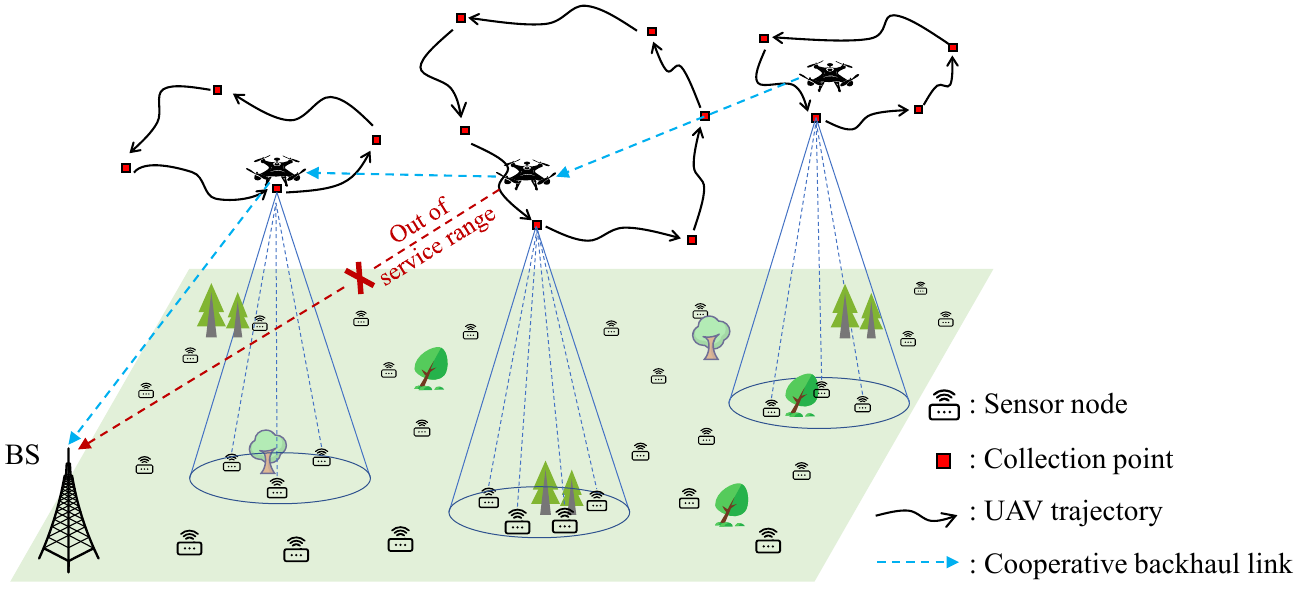}}
    \vspace{-3mm}
    \caption{Multi-UAV collaborative data collection and transmission.}
    \label{system_model}
\end{figure}

As shown in Fig. \ref{system_model}, consider a wireless sensor network deployed in the region of interest, where $N$ SNs, represented as $ \mathcal{S} = \{s_n|1\leq n \leq N\} $, are randomly located.  
Assume that $M$ UAVs, denoted as $ \mathcal{U} = \{u_m|1\leq m\leq M\} $,  are dispatched to collect and transmit SNs' data collaboratively.
The SNs are divided into $K$ non-overlapped clusters, i.e. $\mathcal{S}_k \subseteq \mathcal{S}, 1\leq k\leq K, \bigcup_k \mathcal{S}_k=\mathcal{S}, \mathcal{S}_{k_1} \cap \mathcal{S}_{k_2} = \emptyset, 1\leq k_1 \neq k_2 \leq K$, and collected by the UAVs at $K$ CPs, represented as $ \mathcal{C} = \{c_k|1 \leq k \leq K\} $. 
The UAVs are assumed to fly at a constant altitude $H$ corresponding to authority regulations and safety considerations.
The CPs are at the same altitude as UAVs, when UAV hovers at the CP $c_k$, the data of the SNs in the cluster $\mathcal{S}_k$ will be collected.
Assume that the UAVs utilize orthogonal frequency channels to not interfere with each other.
In the following, the communication channels are first investigated, then the multi-UAV collaboration design is presented, and finally, the optimization problem is formulated.
\vspace{-6mm}

\subsection{Channel Model}
\vspace{-1mm}
The SN-UAV communication channel is assumed to be a probabilistic line-of-sight (LoS) channel considering both the LoS link and non-line-of-sight (NLoS) link. 
The probability of LoS can be approximated to the following function,
\begin{equation}
    P_{\text{LoS}} = \frac{1}{1+a\exp\left(-b\left(\frac{180^\circ}{\pi}\tan^{-1}\left(\frac{H}{r}\right)-a\right)\right)},
\end{equation}
where $a, b$ are the parameters related to the environment, and $r$ is the horizontal distance between the SN and the UAV.
With fixed $H$, the expected path loss of the SN-UAV link can be derived as
\begin{equation}\label{A2G}
    \beta_{\text{G2U}}(r) \! =\! (P_{\text{LoS}} \!+\! P_{\text{NLoS}}\kappa) \beta_0 {d(r)}^{-\alpha},
\end{equation}
where $\beta_0$ denotes the path loss at the reference distance, $d(r) = \sqrt{H^2+r^2}$ is the distance between the SN and the UAV, $\alpha$ is the path loss exponent, $P_{\text{NLoS}} = 1-P_{\text{LoS}}$ is the probability of the NLoS link, and $\kappa < 1$ is the additional attenuation factor due to the NLoS condition. 

Furthermore, the expected signal-to-noise (SNR) received from the SN at the UAV is given by
\begin{equation}
    \gamma_{\text{G2U}}(r) = \frac{P_\text{s} \beta_{\text{G2U}}(r)}{\sigma^2},
\end{equation}
where $P_\text{s}$ is the transmission power of the SN, $\sigma^2$ is the power of channel noise.
To guarantee the successful communication link between the SN and the UAV, the SNR is required to be greater than a pre-specified threshold $\gamma_{\text{G2U}}^{\text{th}}$. 
This requirement can be satisfied by a distance constraint $ r \leq r_{\text{G2U}} $ since $\gamma_{\text{G2U}}(r)$ decreases monotonically as $r$ increases.
Therefore, $r_{\text{G2U}}$ can be obtained by setting $\gamma_{\text{G2U}}(r)=\gamma_{\text{G2U}}^{\text{th}}$, which is named the UAV ground coverage radius hereafter.

Assuming that the UAV collects data at CPs via frequency division multiple access (FDMA) to ensure no interference between SNs \cite{Sabzehali2022optimizing}.
The horizontal coordinate of SN $s_n$ is denoted as $\mathbf{w}_n = (x_n^\text{s},y_n^\text{s})$, the horizontal position of CP $c_k$ is represented as $\mathbf{p}_k = (x_k^\text{c},y_k^\text{c})$. 
Thus, the transmission rate between the SN in $\mathcal{S}_k$ and the UAV at $c_k$ is given by
\begin{equation}
    R_{n,k} = \alpha_n B \log_2(1+\gamma_{\text{G2U}}(\Vert \mathbf{w}_n- \mathbf{p}_k\Vert)), \forall s_n \in \mathcal{S}_k,
\end{equation}
where $\alpha_n$ is the bandwidth allocation ratio of $s_n$ which satisfy $\sum_{s_n \in \mathcal{S}_k}\alpha_n = 1$ and $0 \leq \alpha_n \leq 1, \forall s_n \in \mathcal{S}_k$, and $B$ is the total communication bandwidth.
The number of information bits that need to be collected from the SN $s_n$ is denoted as $Q_n$.
Then the UAV hovering time at $c_k$ for data collection of the SN cluster $\mathcal{S}_k$ can be expressed as $T_k^\text{h} = \max_{s_n \in \mathcal{S}_k} \frac{Q_n}{R_{n,k}}$.

It can be verified that the minimum hovering time is achieved when SNs in $\mathcal{S}_k$ complete data upload at the same time, thus we have $\alpha_n^* = \frac{\rho_n}{\sum_{s_n \in \mathcal{S}_k}\rho_n}, \forall s_n \in \mathcal{S}_k$,
where $\rho_n \triangleq \frac{Q_n}{B \log_2(1+\gamma_{\text{G2U}}(\Vert \mathbf{w}_n- \mathbf{p}_k\Vert))}$. As a result, the UAV minimum hovering time at $c_k$ can be calculated as
\begin{equation}
    \underline{T_k^\text{h}} = \sum_{s_n \in \mathcal{S}_k}\frac{Q_n}{B \log_2(1+\gamma_{\text{G2U}}(\Vert \mathbf{w}_n- \mathbf{p}_k\Vert))}.
\end{equation}

Since UAVs fly over SNs, we assume that their altitudes are high enough to maintain the LoS channel between each other. 
Then the SNR of the UAV-UAV channel can be expressed as 
\begin{equation}
    \gamma_{\text{U2U}} = \frac{P_\text{u} \beta_0}{\sigma^2 d_{\text{u}}^2}, i,j\in[1,M],i \neq j,
\end{equation}
where $P_\text{u}$ is the transmission power of the UAV, and $d_{\text{u}}$ is the distance between two UAVs. 
Assume that the SNR threshold of UAV-UAV link is $\gamma_{\text{U2U}}^{\text{th}}$, then the UAV-UAV coverage radius can be obtained as $r_{\text{U2U}} = \sqrt{\frac{P_\text{u} \beta_0}{\sigma^2 \gamma_{\text{U2U}}^{\text{th}}}}$.

Similar to the SN-UAV channel (c.f. the formula (\ref{A2G})), 
the UAV-BS coverage radius $r_{\text{U2B}}$ can be obtained according to the SNR threshold of the UAV-BS channel $\gamma_{\text{U2B}}^{\text{th}}$.

\subsection{Collaborative Data Collection and Transmission}
Assuming that the BS acts as the central controller with powerful computing ability, intelligently controls the UAVs for collaboration of data collection and transmission, and provides data processing service.
In the mission, all UAVs start at their initial locations, then fly to the CPs for data collection while keeping cooperative backhaul connectivity with the BS for real-time data transmission, and finally, return to their initial positions.
Define a binary variable $a_{m,k}\in \{0,1\}, \forall u_m \in \mathcal{U}, \forall c_k \in \mathcal{C}$, where $a_{m,k} = 1$ indicates that the CP $c_k$ is served by the UAV $u_m$ for the data collection of SN cluster $\mathcal{S}_k$ and $a_{m,k} = 0$ otherwise. 
Since each CP can only be served by one UAV, thus we have
\begin{equation}
    \sum_{m=1}^{M} a_{m,k} = 1, k \in [1,K].
    \label{c1}
\end{equation}

Let $T$ denote the mission completion time, the position of the UAV $u_m$ at time $t$ is represented as $\mathbf{q}_m(t)=(x_m^\text{u}(t), y_m^\text{u}(t)), t \in [0,T]$. 
Introducing the collecting time scheduling variable $\lambda_{m,k}(t) \in \{0,1\}, t \in [0,T]$, where $\lambda_{m,k}(t) = 1$ indicates that the UAV $u_m$ is collecting SNs' data at $c_k$ at the time instant $t$, and $\lambda_{m,k}(t) = 0$ otherwise. 
Since $u_m$ visits $c_k$ only when $a_{m,k}=1$, thus we have,
\begin{equation}
    \lambda_{m,k}(t) \leq a_{m,k}, m \in [1,M], k \in [1,K], t\in[0,T],
    \label{c2}
\end{equation}
\begin{equation}
    \begin{aligned}
    \hspace{-2mm}
        \!-\epsilon \!-\! \phi(1\!-\!&\lambda_{m,k}(t)) \! \leq \! \Vert \mathbf{q}_m(t)\!-\! \mathbf{p}_k \Vert \! \leq \! \epsilon \!+\! \phi(1\!-\!\lambda_{m,k}(t)),\\
        &m \in [1,M], k \in [1,K], t\in[0,T],
    \end{aligned}
    \label{c3}
\end{equation}
where $\epsilon$ is a very small number and $\phi$ is a sufficiently large number, the formula (\ref{c3}) denotes that the UAV $u_m$ is at the CP $c_k$ if $\lambda_{m,k}(t) = 1$.

To guarantee the data collection requirement, the UAV hovering time at CP $c_k$ should satisfy
\begin{equation}
    \int_{0}^{T} \sum_{m=1}^{M} \lambda_{m,k}(t) dt \geq \underline{T_k^\text{h}}, k \in [1,K].
    \label{c4}
\end{equation}

For data transmission, define a set $\mathcal{V}(t)=\{\mathbf{q}_0(t),\mathbf{q}_1(t),\dots,\mathbf{q}_M(t)\}$ that consists of the horizontal positions of the BS and the UAVs at time $t$, where $\mathbf{q}_0(t) = \mathbf{q}_0 = (x^\text{B}, y^\text{B}), t\in [0,T]$ denotes the position of the BS.
Consider a binary variable $z_{ij}(t) \in \{0,1\}$, where $z_{ij}(t) = 1$ if node $\mathbf{q}_i(t)$ is connected to node $\mathbf{q}_j(t)$ at time instant $t$, otherwise $z_{ij}(t) = 0$, $i,j\in[0,M],t\in[0,T]$. According to the SNR conditions, $z_{ij}(t)$ can be activated only when the nodes in $\mathcal{V}(t)$ satisfy the distance constraints,
\begin{equation}
    z_{ij}(t)d_{ij}(t) \leq r_{ij}^\text{th}, i,j \in [0,M], i\neq j, t \in [0,T],
    \label{c5}
\end{equation}
where $r_{ij}^\text{th} = r_{\text{U2U}},\ i,j \in [1,M], i\neq j$ and $r_{i0}^\text{th} = r_{0j}^\text{th} = r_{\text{U2B}},\ i,j \in [1,M]$.

To enable the full-time connectivity between UAVs and BS, there should be at least $M$ connections among $M+1$ nodes in $\mathcal{V}(t)$ and no cycle in any subset of UAVs. Thus, we have 
\begin{align}
    \sum_{i,j \in [0,M], i \neq j} z_{ij}(t) \geq M, t \in [0,T],
    \label{c6}
\end{align}
\begin{equation}
    \sum_{i,j \in [1,M], i \neq j} z_{ij}(t) \! \leq \! \vert \mathcal{G}(t) \vert \!-\! 1,   
    \forall \mathcal{G}(t) \! \subseteq \! \mathcal{V}(t) \! \setminus \! \{\mathbf{q}_0\}, t \! \in \! [0,T], 
    \label{c7}
\end{equation}
where $\mathcal{G}(t)$ is the subset of the UAVs' positions at time $t$, $\vert \mathcal{G}(t) \vert$ denotes the cardinality of the set $\mathcal{G}(t)$.

\subsection{Problem Formulation}
Based on the previous discussions, we present an optimization problem with the objective to minimize the mission completion time of data collection and transmission, i.e.,
\begin{align}
    \text{(P1):}& \min_{\{a_{m,k}\},\{\mathbf{q}_m(t)\},\{\lambda_{m,k}(t)\},\{z_{ij}(t)\},T} T \notag \\ 
    \text{s.t.\ } 
    & (\ref{c1}), (\ref{c2}), (\ref{c3}), (\ref{c4}), (\ref{c5}), (\ref{c6}), (\ref{c7}),\notag \\
    & \Vert \mathbf{q}_i(t)\!-\!\mathbf{q}_j(t) \Vert \! \geq \! D_\text{s}, i,j \! \in \! [1,M], i\neq j, t \! \in \! [0,T], \label{c8}\\
    & \Vert \mathbf{\dot{q}}_m(t) \Vert \leq V_\text{max}, m\in[1,M], t \in [0,T], \label{c9} \\
    & \mathbf{q}_m(0) = \mathbf{q}_m(T), m \in [1,M], \label{c10}
\end{align}
where $\Vert \mathbf{\dot{q}}_m(t) \Vert$ is the instantaneous UAV velocity, $D_\text{s}$ is the safe distance for collision avoidance. 
The constraints in (\ref{c1})-(\ref{c4}) guarantee the data collection requirements of all SNs. 
The constraints in (\ref{c5})-(\ref{c7}) ensure that the UAVs can keep backhaul connectivity with the BS for real-time data transmission at any time.
The minimum distance between UAVs is limited in (14).
The constraint (15) restricts the UAV speed. 
The constraint (16) indicates that the UAVs return to their initial positions after completing the mission.





Note that (P1) is non-convex and involves infinite variables with closely coupled trajectories of UAVs, which is difficult to be optimally solved.
Hence, in the following, we first cluster the SNs and divide the region to determine the location of the CPs and their association with UAVs. 
Then, the formulated problem is simplified and the trajectories of UAVs are optimized in their corresponding regions.

\section{Connectivity-Maintained Multi-UAV Trajectory Optimization}

\subsection{SN Clustering and Region Division}
To cluster the SNs in a load-balance way, we develop an algorithm that can adaptively determine the number of SN clusters while considering the UAV ground coverage and service capacity constraints.
Firstly, we initialize the number of SN clusters as $K=\left \lceil \frac{N}{N_\text{th}} \right \rceil$, where $N_\text{th}$ is the maximum number of SNs that the UAV can serve simultaneously.
Then, the K-means++ algorithm is exploited to cluster the SNs. 
If the number of SNs in a cluster exceeds the capacity $N_\text{th}$ or the distance between SNs and the cluster center exceeds the distance threshold $r_{\text{G2U}}$, the cluster number should be increased and the SNs are re-clustered. 
The details are summarized in \textbf{Algorithm \ref{cluster}}.

To reduce the operation cost of the UAVs, the minimum number of UAVs is used to complete the data collection and transmission in the mission area, which can be obtained as
\begin{equation}
    M = \left \lceil \frac{\max_{1 \leq k \leq K} \| \mathbf{p}_k- \mathbf{q}_0 \|-r_{\text{U2B}}}{r_{\text{U2U}}} \right \rceil + 1.
\end{equation}

To ensure that the UAV can connect to the BS in the worst case, the mission region division relies on the UAV-BS and UAV-UAV coverage radius. 
The first subregion is $\mathcal{R}_1= \{(x,y) \mid \|(x,y)-(x^\text{B},y^\text{B})\|\leq r_{\text{U2B}}, x,y \in \mathbb{R}\}$, and the $m$-th subregion is $\mathcal{R}_m=\{(x,y) \mid r_{\text{U2B}}+(m-2)r_{\text{U2U}} \leq \|(x,y)-(x^\text{B},y^\text{B})\|\leq r_{\text{U2B}}+(m-1)r_{\text{U2U}}, x,y \in \mathbb{R}\}$ for $m \geq 2$.
The CPs in $m$-th subregion are served by the $m$-th UAV, thus the CP-UAV association can be expressed as 
\begin{equation}
    a_{m,k} = \left\{
    \begin{aligned}
        & 1, \ \mathbf{p}_k \in \mathcal{R}_m, \\
        & 0, \ \text{otherwise}.
    \end{aligned}
    \right.
\end{equation}
Under this region division, the UAVs can always be connected to the BS anywhere by adjusting the positions of UAVs.

\begin{algorithm}[t]
    \footnotesize
    \caption{SN clustering and CP location determination}
    \begin{algorithmic}[1]
		\STATE Initialize the UAV ground coverage radius $r_{\text{G2U}}$, the UAV service capacity $N_\text{th}$, the cluster number $K=\left \lceil \frac{N}{N_\text{th}} \right \rceil$.
        \STATE Run K-mean++, get the SN clusters $\mathcal{S}_k$ and the centroids $\mathbf{p}_k$.
        \STATE Calculate $d_\text{max}=\max_{s_n \in \mathcal{S}_k, c_k \in \mathcal{C}} \Vert \mathbf{w_n}-\mathbf{p}_k \Vert $.
        \STATE Calculate $N_\text{max} = \max_{1 \leq k \leq K} \vert \mathcal{S}_k \vert$.
		\IF{$d_\text{max} > r_{\text{G2U}}$ or $N_\text{max} > N_\text{th}$}
		\STATE $K=K+1$.
        \STATE Return to line 2.
        \ENDIF
        \RETURN SN clusters $\mathcal{S}_k (1 \leq k \leq K)$  and CP locations $\mathbf{p}_k (1 \leq k \leq K)$.
    \end{algorithmic}
    \label{cluster}
\end{algorithm}

\vspace{-1mm}
\subsection{Problem Transformation}

\emph{1) Trajectory Descreitization:}
To obtain a more tractable form with a finite number of optimization variables, (P1) can be reformulated by discretizing the trajectories of UAVs under the following connectivity conditions. 

\textbf{Proposition 1:} 
Suppose that the UAV $u_i$ and UAV $u_j$ fly from the initial locations $\mathbf{q}^\text{s}_i$ and $\mathbf{q}^\text{s}_j$ to the final points $\mathbf{q}^\text{e}_i$ and $\mathbf{q}^\text{e}_j$ in the straight lines simultaneously, then $u_i$ and $u_j$ can be connected in the flight if $\max\{\|\mathbf{q}^\text{s}_i-\mathbf{q}^\text{s}_j\|, \|\mathbf{q}^\text{e}_i-\mathbf{q}^\text{e}_j\|\} \leq r_{\text{U2U}}$.

\emph{Proof:} 
Due to page limitation, please refer to Appendix A of the online version in \cite{appendix}.
\hfill $\blacksquare$

Therefore, by applying Proposition 1, the connectivity constraints of UAVs in the whole mission can be expressed as a set of discrete waypoint pairs within the communication range and line segments connected by these points.
With the trajectory discretization, the trajectory of UAV $u_m$ is discretized into $K$ waypoints $\{\mathbf{q}_{m,k}\}, k \in [1,K]$, and $\mathbf{q}_{m,k}$ corresponds to the hovering location of UAV $u_m$ when the data of cluster $\mathcal{S}_k$ is collected. 
Let $\mathbf{I}=(I_1, I_2,\dots, I_K)$ be the serving sequence of the SN clusters.
Thus the trajectory of UAV $u_m$ can be described as follows:  
it starts from its first waypoint $\mathbf{q}_{m,I_1}$, and then flies to the next waypoint $\mathbf{q}_{m,I_2}$ until all waypoints are visited, finally returns to the first waypoint.
Furthermore, with the obtained $\{a_{m,k}\}$, the connectivity constraints in (\ref{c6}) and (\ref{c7}) can be rewritten as 
\begin{align}
    \Vert \mathbf{q}_{1,I_k} - \mathbf{q}_0 \Vert &\leq r_{\text{U2B}}, k \in [1,K],
    \label{c13} \\
    \Vert \mathbf{q}_{m,I_k} - \mathbf{q}_{m-1, I_k} \Vert &\leq r_{\text{U2U}}, k \in [1,K], m \in [2,M],
    \label{c14}
\end{align}
\begin{equation}
    \begin{aligned}
        \frac{\| \mathbf{q}_{m,I_{k+1}}-\mathbf{q}_{m,I_{k}} \|}{v_{m,I_k}} = \frac{\| \mathbf{q}_{m-1,I_{k+1}}-\mathbf{q}_{m-1,I_{k}} \|}{v_{m-1,I_k}}, \\
        k \in [1,K], m \in [2,M],
        \label{c15}
    \end{aligned}
\end{equation}
where $\mathbf{q}_{m,I_{k+1}} \triangleq \mathbf{q}_{m,I_1}$, and $v_{m,I_k}$ denotes the speed of the UAV $u_m$ from $\mathbf{q}_{m,I_k}$ to $\mathbf{q}_{m,I_{k+1}}$. 




\addtolength{\topmargin}{0.1in}%
\emph{2) Completion Time:}
Let $T_m^\text{f}$ and $T_m^\text{h}$ be the total flying time and total hovering time of UAV $u_m$, respectively, then we have 
\begin{equation}
    T_m^\text{f} 
    \!=\! \sum_{k=1}^{K} \frac{\Vert \mathbf{q}_{m,I_{k+1}}\!-\!\mathbf{q}_{m,I_{k}}\Vert }{v_{m,I_k}}
    \!=\! \sum_{k=1}^{K} \frac{\max_m \|\mathbf{q}_{m,I_{k+1}}\!-\!\mathbf{q}_{m,I_{k}} \|}{V_\text{max}}.
\end{equation}

Denote $\tau_{m,I_k}$ as the hovering time for UAV $u_m$ at the location $\mathbf{q}_{m,I_k}$, Then we have
$T_m^\text{h} = \sum_{k=1}^{K} \tau_{m,I_k}.$

Since the UAV needs to hover for enough time for data collection of SNs, the data collection constraint in (\ref{c4}) can be rewritten as 
\begin{equation}
    \tau_{m,I_k} \geq a_{m, I_k} \underline{T_{I_k}^\text{h}}, m \in [1,M], k \in [1,K].
    \label{c16}
\end{equation}

It can be verified that at the optimal solution, all the UAVs have the same total flying time and total hovering time, i.e. $T_i^\text{f} = T_j^\text{f}$ and $T_i^\text{h} = T_j^\text{h}$, $i,j\in[1,M], i \neq j$, thus the mission completion time can be represented as $T = T_m^\text{f}+T_m^\text{h}$.

\textbf{Lemma 1:} 
The total hovering time of the UAV can be decreased if multiple UAVs hover over their allocated CPs and perform data collection simultaneously with maintained backhaul connectivity link. 

\emph{Proof:} 
Please refer to Appendix B in \cite{appendix}.
\hfill $\blacksquare$

Define $\mathcal{P}_{I_k} = \{\mathbf{q}_{m,I_k}\}_{m=1}^{M} \cap \{ \mathbf{p}_{I_k}\}_{k=1}^{K}$,
for a group of positions of UAVs for collecting data, at least one UAV is hovering over a CP, thus we have
\begin{equation}
    |\mathcal{P}_{I_k}| \geq 1, k\in[1,K].
    \label{c17}
\end{equation}
Inspired by Lemma 1, the constraint (\ref{c16}) can be rewritten as
\begin{equation}
    \sum_{i \in \mathcal{A}_{I_k}} \tau_{m,i} \geq \max_{i \in \mathcal{A}_{I_k}} \underline{T_{i}^\text{h}}, m \in [1,M], k\in[1,K],
    \label{c18}
\end{equation}
where $\mathcal{A}_{I_k}$ is the index set of $\mathcal{P}_{I_k}$ in $\{ \mathbf{p}_{I_k}\}_{k=1}^{K}$.

With the above manipulations, the problem (P1) reduces to finding the optimal UAV hovering locations $\{\mathbf{q}_{m,I_k}\}$, hovering time $\{\tau_{m,I_k}\}$, and the CP serving sequence $\mathbf{I}$, i.e.,
\begin{align}
    \text{(P2):}&\min_{\{\mathbf{q}_{m,I_k}\}, \{\tau_{m,I_k}\}, \mathbf{I}} T \notag \\
    \text{s.t.\ }
    & (\ref{c13}), (\ref{c14}), (\ref{c15}), (\ref{c17}), (\ref{c18}), \notag \\
    & -\! \epsilon \!-\! \phi(1\!-\! a_{m, I_k}) \! \leq \! \Vert \mathbf{q}_{m,I_k} \!- \! \mathbf{p}_{I_k} \Vert \!\leq\! \epsilon \!+\! \phi(1 \!-\!a_{m, I_k}),\notag \\
    & k \in [1,K], m \in [1,M],\\
    & \Vert \mathbf{q}_{m,I_k} \! -\! \mathbf{q}_{m-1, I_k} \Vert \! \geq \! D_\text{s}, k \in [1,K], m \in [2,M].
\end{align}
Although the problem is complicated since the optimization of the CP serving sequence has a combinatorial nature, the traveling salesman problem (TSP) path of each UAV holds promise for efficient trajectory optimization.

\subsection{Point-Matching-Based Trajectory Planning}
With the relaxation of the connectivity constraints of UAVs in the mission, it can be found that the optimal flight path of each UAV $u_m$ completes the data collection task in its allocated area is the TSP path connecting the allocated CPs.
Denote the set of CPs associated with UAV $u_m$ is $\mathcal{C}_{m}=\{c_{k}|a_{m,k}=1,1\leq k\leq k\}$, the TSP path of UAV $u_m$ for visiting the CPs in $\mathcal{C}_m$ is represented as $\Gamma_{m}^{\text{TSP}}=(\pi_{m,1},\cdots,\pi_{m,|\mathcal{C}_m|})$.
For notational convenience, define $\mathbf{p}_{\pi_{m,1}} = \mathbf{p}_{\pi_{m,|\mathcal{C}_m|+1}}$, then the total flight distance of the TSP path can be obtained as
$d_{m}^{\text{TSP}}=\sum_{l=1}^{|\mathcal{C}_m|}\|\mathbf{p}_{\pi_{m,l+1}}-\mathbf{p}_{\pi_{m,l}}\|.$
Furthermore, the mission completion time of UAV $u_m$ can be expressed as
$T_{m}^{\text{TSP}}=\frac{d_{m}^{\text{TSP}}}{V_\text{max}}+\sum_{c_{k}\in \mathcal{C}_{m}}\underline{T_{k}^{h}}.$
Therefore, the lower bound of the mission completion time can be obtained as
\begin{equation}
    \underline{T} = \max_m T_m^{\text{TSP}}.
    \label{lower bound}
\end{equation}

However, when the connectivity constraints are introduced, UAVs may need to adjust their flight trajectory to guarantee the connection with BS, which may lead to a longer flight distance and increase the time overhead.
Denote the index of the slowest UAV is $A = \arg\max_{m} T_m^{\text{TSP}}$.
One effective solution is that under the condition that $u_{A}$ maintains the TSP path, the flight path of the UAV in the adjacent area ($u_{A-1}$ or $u_{A+1}$) is adjusted to keep it connected to $u_{A}$, so as to ensure the minimum increase to the lower bound of the total mission completion time.
Next, take the UAV $u_{A-1}$ as an example, we proposed an efficient algorithm to plan the path of UAV $u_{A-1}$ based on point matching. 
The algorithm consists of four steps:

\textbf{Step 1}: \emph{Connectable CP Set Mapping.}
Define $c_{k}$ and $c_{\hat{k}}$ are the connectable CPs if they satisfy $\|\mathbf{p}_{k}-\mathbf{p}_{\hat{k}}\| \leq r_{\text{U2U}}$, where $c_{k} \in \mathcal{C}_{A-1}$ and $c_{\hat{k}} \in \mathcal{C}_{A}$.
Therefore, for any CPs in $\mathcal{C}_{A-1}$, the corresponding connectable CP set in $\mathcal{C}_{A}$ can be presented as 
\begin{equation}
    \mathcal{B}_{k} = \{ c_{\hat{k}} \mid \|\mathbf{p}_{\hat{k}}-\mathbf{p}_{k}\| \leq r_{\text{U2U}}, c_{\hat{k}} \in \mathcal{C}_{A} \}, \forall c_{k} \in \mathcal{C}_{A-1}.
\end{equation}

\textbf{Step 2}: \emph{CP Pair Matching.}
Denote the CP that matches $c_{k}$ is $c_{k'}$, in order not to increase the extra hovering time of the UAV $u_{A}$, the CP matching $c_{k}$ in $\mathcal{B}_{k}$ should be satisfied
\begin{equation}
    \underline{T_{k}^\text{h}} \leq \underline{T_{k'}^\text{h}}.
\end{equation}
Furthermore, in order not to increase the extra flight time of the UAV $u_{A}$, the two consecutive CPs in $\mathcal{C}_{A-1}$ and their matched CPs in $\mathcal{C}_{A}$ should be satisfied
\begin{equation}
    \|\mathbf{p}_{k_1}-\mathbf{p}_{k_2}\| \leq d^{\text{TSP}}(c_{k'_1} \rightarrow c_{k'_2}),
\end{equation}
where $d^{\text{TSP}}(c_{k'_1} \rightarrow c_{k'_2})$ denotes the distance of the TSP path from $c_{k'_1}$ to $c_{k'_2}$.

\textbf{Step 3}: \emph{Unconnectable CP Corresponding Waypoint Generation.}
If there exists $\mathcal{B}_{k}=\emptyset$, it means that there is no CP in $\mathcal{C}_{A}$ can be connected to $c_{k}$ in $\mathcal{C}_{A-1}$.
Therefore, the new waypoint should be generated in the region $\mathcal{R}_A$ to connect with $c_{k}$ when the UAV $u_{A-1}$ is hovering over $c_{k}$ for collecting data.
To make the flight distance of UAV $u_{A}$ as small as possible, the new waypoint $\mathbf{q}_{k'}$ can be obtained by solving
\begin{align}
    \text{(P3):}&\min_{\mathbf{q}_{k'},l}  
    \|\mathbf{p}_{\pi_{A,l}} \!-\! \mathbf{q}_{k'}\| \!+\!
    \|\mathbf{q}_{k'} \!-\! \mathbf{p}_{\pi_{A,l+1}}\| \!-\! 
    \|\mathbf{p}_{\pi_{A,l}} \!-\! \mathbf{p}_{\pi_{A,l+1}}\|  \notag \\
    \text{s.t.\ }
    & D_\text{s} \leq \|\mathbf{q}_{k'}-\mathbf{p}_{k}\| \leq r_{\text{U2U}}, \mathbf{q}_{k'} \in \mathcal{R}_{A}, \\
    & 1 \leq l \leq |\mathcal{C}_{A}|-1,
    \label{new point}
\end{align}
where $\mathbf{q}_{k'}$ is the coordinates of the new waypoints connected to $c_{k}$. 
Denote the coordinates of $\mathbf{p}_{k}$, $\mathbf{p}_{\pi_{A,l}}$ and $\mathbf{p}_{\pi_{A,l+1}}$ as $(x_0, y_0)$, $(x_1,y_1)$ and $(x_2,y_2)$, with fixed $l$, the optimal new waypoints can be expressed as 
$\mathbf{q}^*_{k'} = \left( \frac{Xr_{\text{U2U}}}{\sqrt{X^2+Y^2}}+x_0,\frac{Yr_{\text{U2U}}}{\sqrt{X^2+Y^2}}+y_0 \right),$
where $X=x_1+x_2$ and $Y=y_1+y_2$.
Finally, calculate and compare the objective values with different $l$, the optimal solution can be obtained.
Then, the corresponding hovering time at $\mathbf{q}_{k'}$ is $\underline{T_{k'}^\text{h}}=\underline{T_{k}^\text{h}}$.

\textbf{Step 4}: \emph{Path Planning.}
The trajectory of UAV $u_{A}$ can be obtained by inserting the generated waypoints added in Step 2 into the original TSP path of UAV $u_{A}$ in subregion $\mathcal{R}_{A}$.
The trajectory of UAV $u_{A-1}$ can be obtained by inserting extra waypoints (marked in grey) which should be connected to the unmatched CPs in $\mathcal{C}_{A}$ into the TSP path of UAV $u_{A-1}$ in subregion $\mathcal{R}_{A-1}$.
Similar to step 2, each extra waypoint in subregion $\mathcal{R}_{A-1}$ should be satisfied
\begin{align}
    D_\text{s} \leq \|\mathbf{q}_{k}-\mathbf{p}_{k'}\| \leq r_{\text{U2U}},\\
    \|\mathbf{q}_{k_1}-\mathbf{q}_{k}\| \leq \|\mathbf{p}_{k'_1}-\mathbf{p}_{k'}\|,\\
    \|\mathbf{q}_{k_2}-\mathbf{q}_{k}\| \leq \|\mathbf{p}_{k'_2}-\mathbf{p}_{k'}\|,
\end{align}
where $\mathbf{q}_{k}$ denotes one of the the extra waypoint for UAV $u_{A-1}$, $\mathbf{q}_{k_1}$ and $\mathbf{q}_{k_2}$ are the waypoints are adjacent to $\mathbf{q}_{k}$ on the path of UAV $u_{A-1}$.
The details of the overall algorithm are summarized in \textbf{Algorithm \ref{PMTPA}}.

\begin{algorithm}[t]
    \footnotesize
    \caption{Point-matching-based trajectory planning}
    \begin{algorithmic}[1]
        \STATE \textbf{Input}: CP locations $\mathbf{p}_k(1 \leq k \leq K)$, regions $\mathcal{R}_m(1 \leq m \leq M)$, CP set associated with UAV $\mathcal{C}_m(1 \leq m \leq M)$.
		\STATE Initialize each UAV path with the TSP path for visiting the CPs in its associated CP set, $A = A' = \arg\max_{m} T_m^{\text{TSP}}$.
        \REPEAT
        \STATE \textbf{Step 1}: Calculate the connectable CP set for each CP in $\mathcal{C}_{A-1}$.
        \STATE \textbf{Step 2}: Match the CP pairs in $\mathcal{C}_{A-1}$ and $\mathcal{C}_{A}$ through (30) and (31).
        \STATE \textbf{Step 3}: Calculate the optimal waypoint for unconnectable CP in $\mathcal{C}_{A-1}$ by solving (P3).
        \STATE \textbf{Step 4}: Insert the waypoints into the TSP path through (34)-(36).
        \STATE Execute Step 1-Step 4 for $\mathcal{C}_{A'+1}$.
        \STATE Update $A = A-1, A'=A'+1$.
        \UNTIL{$A=1, A'=M$}.
        \STATE \textbf{Output}: Trajectories of all the UAVs.
    \end{algorithmic}
    \label{PMTPA}
\end{algorithm}

\section{Simulation and Results Analysis}
In this section, numerical results are provided to validate the proposed algorithm. We consider a geographical region of size $8\  \text{km} \times 8 \ \text{km}$, where 1000 SNs are deployed randomly and the BS is located at the lower left corner of the area.
The parameters in the suburban environment for the probabilistic LoS channel model are set as $a=4.88$, $b=0.43$, $\kappa=0.2$, $\alpha=2$.
We set $H$ = 100 m, $H^\text{B}= 20$ m, $V_\text{max}$ = 30 m/s, $D_s$ = 30 m $N_\text{th}$ = 60.
The communication-related parameters are set as $B$ = 2 MHz, $P_\text{u}$ = 0.1 W, $P_\text{s}$ = 0.05 W, $\sigma^2$ = -110 dBm, $\gamma_{\text{G2U}}^{\text{th}}$ = 20 dB, $\gamma_{\text{U2U}}^{\text{th}}$ = 19.5 dB, $\gamma_{\text{U2B}}^{\text{th}}$ = 13 dB.
We assume that all SNs have an identical amount of data, i.e., $Q_n=Q=10\  \text{Mbits}, \forall s_n \in\mathcal{S}$. 

\begin{figure*}[t]
    \centering
    \subfloat[SN clusters and region division.]{\includegraphics[width=1.6in]{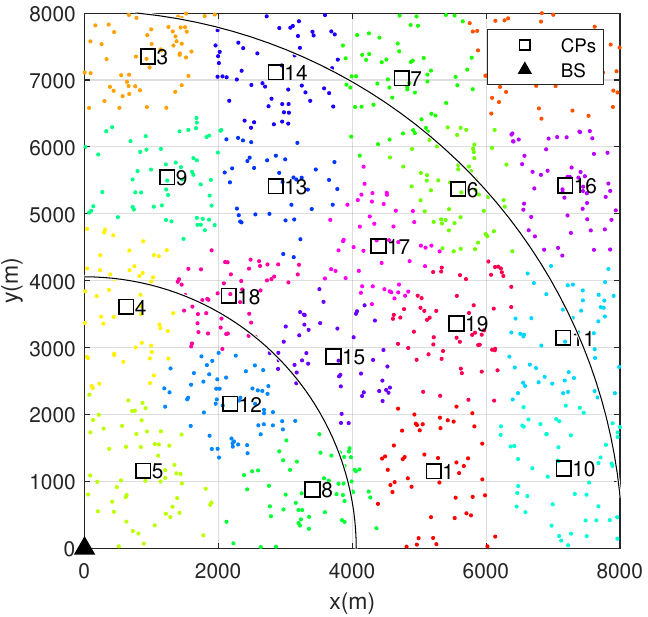}} 
    \subfloat[TTP algorithm (1664.4s).]{\includegraphics[width=1.6in]{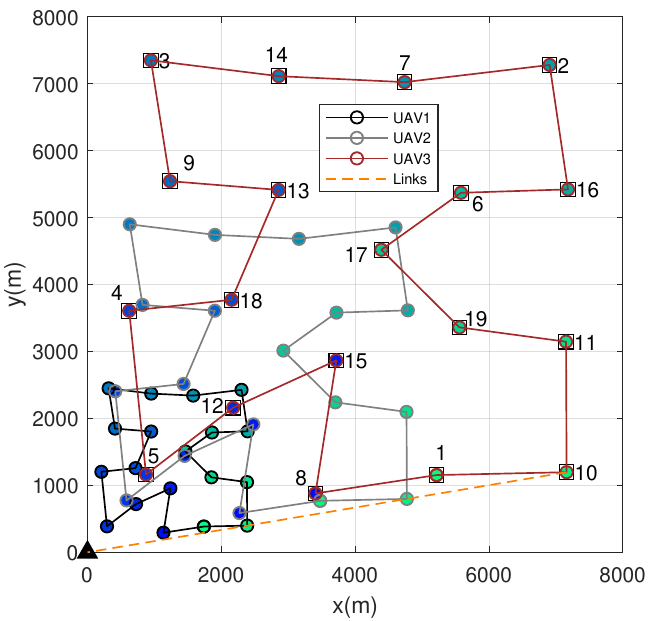}} 
    \subfloat[CSTP algorithm (1564.6s).]{\includegraphics[width=1.6in]{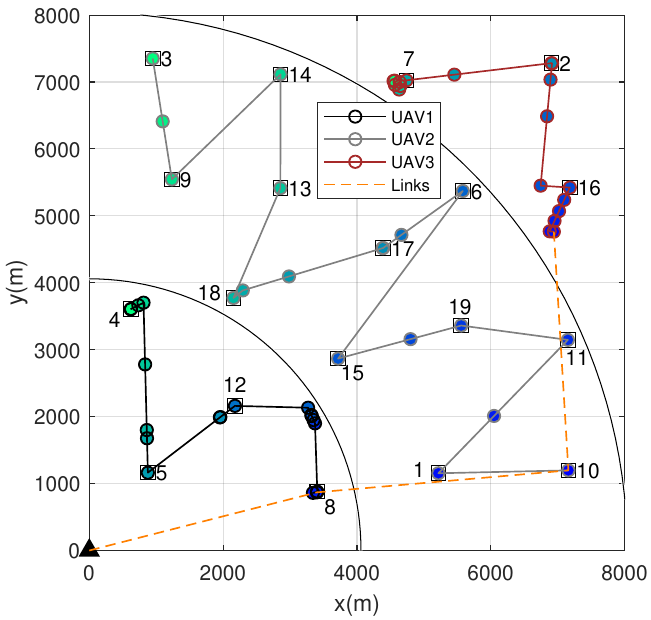}} 
    \subfloat[PMTP algorithm (1068.9s).]{\includegraphics[width=1.6in]{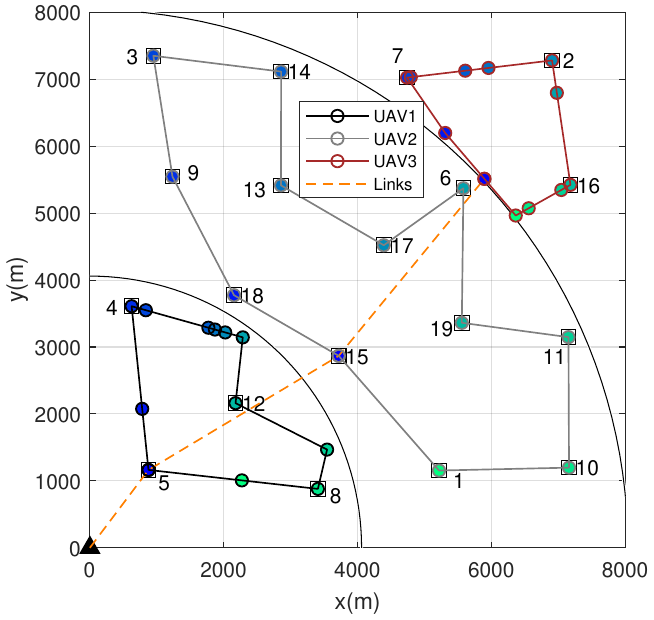}}
    \caption{Performance comparison of different algorithms.}
    \label{UAV trajectory with different algorithms}
\end{figure*}

\begin{figure}[t]
    \centering
    \subfloat[]{\includegraphics[width=1.5in]{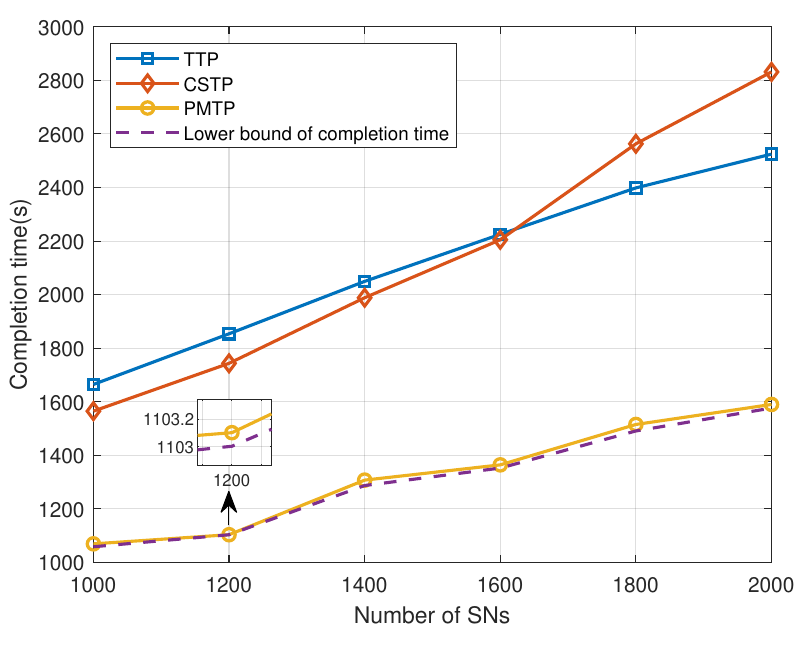}}
    \subfloat[]{\includegraphics[width=1.5in]{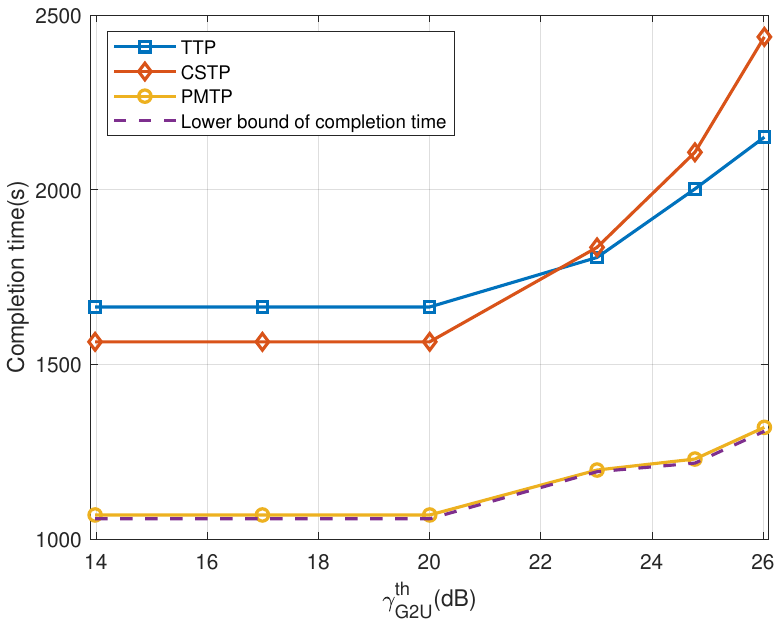}}
    \caption{Completion time versus the number of SNs or $\gamma_{\text{G2U}}^{\text{th}}$}
    \label{time vs SN number or SNR_G2U}
\end{figure}

For evaluation, the proposed algorithm is compared with the following two benchmarks: TSP-based trajectory planning (TTP) and circular-searching-based trajectory planning (CSTP). 
TTP uses one UAV to complete the data collection of all CPs with the TSP path. While the other UAVs, as relays, are always evenly distributed on the line between the collecting UAV and the ground BS during the flight to ensure the connection.
CSTP uses multiple UAVs to collect data in different subregions through circular scanning. The CP serving sequence is determined by the angle between CP and BS, and CP with a small angle is collected first. When a UAV is collecting data at the CP, the other UAVs are hovering in their subregions with maintained connectivity.

Fig. \ref{UAV trajectory with different algorithms}(a) shows the SNs are divided into 19 clusters with approximately equal load, and the mission region is divided into three subregions.
Fig. \ref{UAV trajectory with different algorithms}(b)-(d) shows the UAV trajectories obtained by different algorithms. 
Specifically, the completion time of the proposed algorithm is 35.8\% and 31.7\% lower than that of the benchmarks, respectively.
Compared with the TTP, the CSTP reduces the flight distance through collaborative data collection in different subregions, so the flight time is shorter.
The UAVs in the proposed algorithm can collect multiple CPs at the same time, thus the collection time is reduced, on the other hand, the UAVs' paths are optimized based on the TSP paths in subregions, which leads to shorter flight time, resulting in the completion time decreases significantly.

As shown in Fig. \ref{time vs SN number or SNR_G2U}(a), the completion time gradually increases as the number of SNs increases. 
Specifically, the completion time of the proposed algorithm is on average 37.5\% and 38.4\% lower than that of the two benchmarks, respectively.
When the number of SNs is larger, the completion of the CSTP will exceed that of the TTP, the main reason is that the CP serving sequence through circular scanning will lead to longer flight distance caused by the fluctuating trajectory with the increased number of CPs.
In addition, it can be seen that the completion time achieved by the proposed algorithm is very close to the lower bound (c.f. formula (\ref{lower bound})), and the average gap is about 1.0\%.
In particular, when the number of SNs is 1200, the difference between the completion time of the proposed algorithm and the lower bound is 0.1\%.

Fig. \ref{time vs SN number or SNR_G2U}(b) shows the completion time for different SNR thresholds of SN-UAV link. It can be seen that with the increase in $\gamma_{\text{G2U}}^{\text{th}}$, the completion time also increases.
The main reason is that the increase in $\gamma_{\text{G2U}}^{\text{th}}$ leads to a smaller UAV ground coverage radius, and the SNs need to be divided into more clusters, corresponding to more CPs, which increases the flight distance of the UAV and leads to a longer completion time.
However, when $\gamma_{\text{G2U}}^{\text{th}}$ decreases, although the UAV ground coverage radius increases, due to the limitation of UAV service capacity, the number of SN clusters cannot be further reduced, so the completion time falls into convergence.

\section{Conclusion}
This paper investigated the multi-UAV-enabled real-time data collection and transmission of large-scale sensor networks.
We proposed a multi-UAV collaborative scheme for seamless data collection and transmission, and formulated an optimization problem aiming at minimizing the completion time by optimizing the trajectories, task allocation, collection time scheduling, and transmission topology of UAVs.
The formulated problem was converted into a tractable form by SN clustering, region division, connectivity condition derivation, and trajectory discretization. 
Then, we propose a point-matching-based trajectory planning algorithm to solve the problem efficiently.
The simulation results show that the proposed scheme achieves significant performance gains over the two benchmarks.
\vspace{-3mm}
\bibliographystyle{IEEEtran}
\bibliography{IEEEabrv,reference}

\end{document}